\documentclass[aps,prd,showpacs]{revtex4}
\usepackage{graphicx}
\usepackage{epsfig}
\usepackage{psfig}
\usepackage{revsymb}
\begin{document}
\renewcommand{\thesection}{\arabic{section}}
\renewcommand{\thesubsection}{\arabic{subsection}}

\title{Neutron Star Equation of State and The Possibility of Complex 
Self-Energy in Landau Theory of Fermi Liquid in Presence of Strong Quantizing 
Magnetic Field}
\author{Soma Mandal$^{a)}$, Roni Saha$^{a)}$, Sutapa Ghosh$^{a)}$
and Somenath Chakrabarty$^{b)}$ }
\affiliation{
$^{a)}$Department of Physics, University of Kalyani, Kalyani 741 235, 
West Bengal, India\\
$^{b)}$Department of Physics, Visva-Bharati, Santiniketan 731 235, 
West Bengal, India, E-mail:somenath@vbphysics.net.in}

\begin{abstract}
Using the relativistic version of Landau theory of Fermi liquid with 
$\sigma-\omega$ and $\rho$ mesons exchange, we have obtained an equation 
state for dense neutron star matter in presence of strong quantizing magnetic 
field.  It is found that in this scenario the self-energies of both neutron 
and proton components of dense neutron star matter become complex under certain
physical conditions. To be more specific, it is observed that in the exchange 
diagrams of $\sigma$, $\omega$ and neutral $\rho$ transfer processes 
and in the direct interaction diagram with $\rho_{\pm}$ transfer reactions, the 
nucleon self-energies become complex in nature.
\end{abstract}
\maketitle
\section{Introduction}
With the recent discoveries of a few magnetars \cite{R1,R2,R3,R4}, the study 
of dense neutron star matter in presence of strong quantizing magnetic field 
has gained a new life. These strange stellar objects are supposed to be 
strongly magnetized, relatively young (age $\sim10^4$yrs.) neutron stars and 
are also believed to be the possible sources of anomalous X-ray pulses and 
soft gamma ray emissions (AXP and SGR). As surface magnetic fields of these 
high magnetically-powered neutron stars are observed to be $\geq10^{14}-
10^{15}$G, then a theoretical estimation from virial theorem 
predicts that the field at the core region may be $\sim10^{18}$G. Now, if the 
magnetic fields are really so strong, particularly at the central region, they 
must affect most of the physical properties of those strange stellar objects. 
Also, different physical processes taking place at the core region should get 
modified significantly.

The strong magnetic field at the core region affects the equation of state of 
the neutron star matter and as a consequence, some of the gross properties, 
e.g. mass-radius relation, moment of inertia, rotational frequency etc. of 
neutron stars should change to a great extent \cite{R5,R6,R7,R8}.

In the case of compact neutron stars, the possibility of first order phase 
transition to quark matter at the core region is found to be completely 
forbidden if the magnetic field strength exceeds $10^{15}$G. However, a 
second order phase transition is allowed, provided the field strength is 
$<10^{20}$G \cite{R9,R10}.

We have noticed that the elementary processes, e.g. weak and 
electromagnetic reactions and decays are also strongly influenced by such 
intense magnetic field. As the cooling of neutron stars are mainly controlled 
by neutrino emissions through weak processes (URCA or modified URCA), 
the presence of strong quantizing magnetic field should significantly affect 
the thermal evolution of neutron stars. Further, the rates of the 
electromagnetic processes in presence of strong quantizing magnetic field 
should modify significantly the transport properties of dense stellar matter, 
in particular for electron gas present at the core region of the neutron stars 
\cite{RY1,RY2,RY3,R11,R12}. We have noticed that the electron gas behaves 
like a neutral super fluid if the magnetic field strength exceeds $10^{16}$G 
\cite{R13}.

It has also been observed that chiral symmetry gets violated in presence of 
strong magnetic field. The strong quantizing magnetic field acts like a 
catalyst to generate the mass dynamically. A lot of investigations have
already been done in this field \cite{R14,R15,R16,R17,R18,R19,R20,R21,SCH}.

In the general relativistic model some investigations have also been done to
study the stability of neutron stars in presence of strong magnetic field.
It is found that there are possibilities of geometrical deformations from 
the usual spherical shapes and in the extreme case of ultra strong magnetic 
field, they become either black disks or black strings \cite{R22,R23,R24}.

In one of our recent studies, we have developed a relativistic formalism of 
Landau theory of Fermi liquid for dense neutron star matter in presence of 
strong quantizing magnetic field with $\sigma-\omega$ meson exchange 
\cite{R25}. In the present article we shall re-formulate our previous work as 
cited above, incorporating $\rho$-meson (both charged and neutral components) 
exchange interactions. To obtain the self-energies for both neutron and proton 
components from the Landau quasi-particle interaction functions considering 
nucleonic interactions via the exchange of $\rho_\pm$ mesons, we have also 
included the effect of strong quantizing magnetic field on charged meson 
propagators.

To evaluate the self-energy of Landau quasi-particles, we have considered a 
typical many body fermionic system (in this case it is neutron star matter) in 
presence of strong quantizing magnetic field at zero temperature, interacting 
via scalar, vector and iso-vector meson fields (as mentioned above). This 
formalism is also useful to obtain the quasi-particle energy of other fermionic
many body system, e.g., electron gas or even for exotic quark matter in 
presence of strong magnetic field. The formalism is also applicable to study 
the magnetic properties and the non-equilibrium properties (e.g., the transport
properties) of dense stellar plasma of fermions. Now similar to the 
conventional non-relativistic model of Landau theory of Fermi liquid, this 
modified version also deals with normal Fermi liquid and is applicable for the 
low-lying excited states of the system, which consists of a superposition of 
quasi-particle excitation close to the Fermi surface. The formalism we have 
developed in this article is basically a re-formulation of our previous work 
\cite{R25} and an extension of a very old paper by Baym and Chin \cite{R26}. 
Of course, in the present investigation, we have found a new and interesting 
result, the complex nature of nucleonic self energies in the mean field
approximation.

The paper is arranged in the following manner. In the next section we are 
giving a general overview of the behavior of a fermionic system in a strong
quantizing magnetic field. In Section 2, we shall consider a typical 
dense neutron star matter in presence of strong quantizing magnetic
field interacting via $\sigma-\omega-\rho$ mesons exchange and develop a 
formalism to obtain the equation of state for dense neutron star matter in 
presence of strong quantizing magnetic fields. In the case of $\rho_1$ and 
$\rho_2$ (or $\rho_\pm$, the positive and negative components respectively) 
exchange interactions, we shall incorporate the effect of strong quantizing 
magnetic field on their propagators. In Section 3, we have concluded the 
results and discussed the future perspective of this work.

\section{Basic Formalism}
\subsection{Charged Fermion in Strong Magnetic Field}
In the case of a  typical many body fermionic system, let us consider the 
$i$th type fermion of charge $q_i$ and mass $m_i$ (of course, in the case of 
neutron component of neutron star matter, since $q_i=0$, then one has to use 
the original work by Baym and Chin and the same formalism is also applicable 
if the magnetic field strength is less than the quantum critical value). We 
assume the gauge $A^\mu \equiv(0,0,xB_m,0)$, so that the constant magnetic 
field B$_m$ is along $z$-axis. If the magnetic field strength $B_m$ exceeds 
the quantum critical value $B_m^{(c)(i)}=m^2_i/q_i$, the Landau levels 
for the $i$th charged component will be populated and the matter is
called the Landau diamagnetic system. In the relativistic region the
charged matter shows Landau diamagnetic behavior, if the cyclotron
quantum of the charged components exceeds the corresponding rest mass
energy or equivalently the de Broglie wave length exceeds the Larmor radius 
of the charged particle. Under such circumstances the phase space integral
will get modified and is given by
\begin{eqnarray}
\frac{g_i}{(2\pi)^3}\int d^3pF(p)&=&\frac{g_i}{(2\pi)^3}\int dp_zd^2p_{\perp}
F(p)\nonumber\\
&\longrightarrow & \frac{g_iq_iB_m}{(2\pi)^2}\sum_{\nu=0}^\infty
(2-\delta_{\nu_0})\int_{-\infty}^{+\infty}dp_{z}F(p_z,\nu)
\end{eqnarray} 
where $g_i$ is the degeneracy, $\nu=0,1,2,...$ is the Landau quantum number and 
$F$ is a function of particle momentum. At zero temperature, the above modified
form reduces to
\begin{equation}
\longrightarrow
\frac{g_iq_iB_m}{(2\pi)^2}\sum_{\nu=0}^{[\nu_{max}^{(i)}]}
(2-\delta_{\nu_0})\int_{-p_{F_i}}^{+p_{F_i}}dp_{z}F(p_z,\nu)
\end{equation} 
where   
\begin{eqnarray}
[\nu_{max}^{(i)}]=\frac{(\mu_{i}^{2}-m_{i}^{2})}{2q_{i}B_{m}}
\end{eqnarray} 
is an integer but less than the actual value of right hand side, $\mu_i$
is the chemical potential of the $i$th charged component and $p_{F_i}$
is the corresponding Fermi momentum. Now it can very easily be shown from 
the solution of Dirac equation in presence of such strong magnetic field 
that the zeroth Landau level is singly degenerate, whereas all other levels 
are doubly degenerate (which is incorporated in eqn.(1) by introducing the 
factor $2-\delta_{\nu 0}$). Further, the energy eigen value is given by
\begin{eqnarray}
E_\nu=(p_{z}^{2}+m_{i}^{2}+2\nu q_{i}B_{m})^{1/2}=(p_{z}^{2}+m_{i}^{2}+
p_{\perp,i}^{2})^{1/2}
\end{eqnarray}
and therefore the chemical potential is simply given by
\begin{eqnarray} 
\mu_{i}=(p_{F_i}^{2}+m_{i}^{2}+2\nu q_{i}B_{m})^{1/2}
\end{eqnarray}
with $p_{\perp,i}=(2\nu q_{i}B_{m})^{1/2}$ the orthogonal component of 
charged particle momentum (in the $x-y$ plane), which varies in a discrete 
manner, hence the name, Landau quantization.

In presence of strong quantizing magnetic field as discussed above, it can 
very easily be shown that in standard notation, the modified form of Dirac 
spinors for the charged components are given by
\begin{eqnarray}
\psi=\frac{1}{(L_{y}L_{z})^{1/2}}\exp(-iE_{\nu}t+ip_{y}y+ip_{z}z)
u^{\uparrow \downarrow}
\end{eqnarray} 
where
\begin{equation} 
u^{\uparrow}=\frac{1}{[E_{\nu}(E_{\nu}+m)]^{1/2}}
\left( \begin{array}{c}
(E_{\nu}+m)I_{\nu;p_y}(x)\\0\\p_zI_{\nu;p_y}(x)\\
-i(2\nu q_iB_m)^{1/2}I_{\nu-1;p_y}(x)
\end{array} \right)  
\end{equation}
and
\begin{eqnarray}
u^{\downarrow}=\frac{1}{[E_{\nu}(E_{\nu}+m)]^{1/2}}
\left( \begin{array}{c}
0\\(E_{\nu}+m)I_{\nu-1;p_y}(x)\\i(2\nu q_iB_m)^{1/2}I_{\nu;p_y}(x)\\
-p_zI_{\nu -1;p_y}(x)
\end{array} \right) 
\end{eqnarray}
where the symbols $\uparrow$ and $\downarrow$ indicate the up and the down 
spin states respectively and
\begin{eqnarray}
I_\nu=\left (\frac{qB_m}{\pi}\right )^{1/4}\frac{1}{(\nu !)^{1/2}}2^{-\nu/2}
\exp \left [{-\frac{1}{2}qB_m\left (x-\frac{p_y}{qB_m} \right )^2}\right 
]\nonumber \\
H_\nu \left [(qB_m)^{1/2}\left (x-\frac{p_y}{qB_m} \right) \right ]
\end{eqnarray}
with $H_\nu$ the well-known Hermite polynomial of order $\nu$ and
$L_y,~~L_z$ are length scales along $y$ and $z$ directions respectively.

Now for a dense relativistic many body system of fermions interacting via 
scalar, vector and iso-vector fields, represented by $\phi,V^{\mu}\;{\rm and} \;
\vec{\rho}^{\mu}$ with masses $\zeta, \eta\;{\rm and}\;\theta$
respectively, the basic Lagrangian density is given by
\begin{eqnarray}
{\cal{L}}&=& \frac{i}{2} \bar{\psi} \gamma^{\mu} 
(\vec
D_\mu-D^{\!\!\!\!\!^{^\leftarrow}}_\mu)
\psi-m\bar{\psi}\psi-g_v\bar{\psi}\gamma^{\mu}V_{\mu}\psi-
g_{s}\bar{\psi}\phi\psi-g_{\rho}\bar{\psi}\gamma^\mu\vec{\rho_{\mu}}.
\frac{\vec{\tau}}{2}\psi \nonumber\\&&+
{\cal{L}}_V+{\cal{L}}_S+{\cal{L}}_{I-V}+{\cal{L}}_{EM}
\end{eqnarray}
where $D^{\mu}=\partial^{\mu}+iqA^{\mu},~g_v,~g_s$ and $g_{\rho}$ are the 
coupling constants corresponding to vector,
scalar and iso-vector interactions; ${\cal{L}}_V,~{\cal L}_S,~{\cal L}_{I-V}$
are the free Lagrangian densities for vector, scalar and iso-vector field 
respectively and ${\cal L}_{EM}$ is the Lagrangian density electromagnetic 
part corresponding to constant external magnetic field $B_m$ along $z$-axis.
Here the iso-vector fields behaves like a four vector in configuration space 
(represented by ${\mu}=0,1,2,3$) but like a three vector in iso-spin space
(represented by the symbol $\rightarrow$).

Now in the Landau theory of Fermi liquid, to evaluate various physical
quantities of the system, it is a common practise to switch on the 
interactions adiabatically. The normal Fermi liquid 
system will then evolve from non-interacting states to interacting states 
and there will be a one-to-one correspondence between the dressed or 
quasi-particle states of the interacting system to the bare non
interacting states.
To obtain $f$, the Landau Fermi liquid interaction function, also known 
as the quasi-particle interaction function, we have considered both direct 
and exchange type interactions. The exchanged bosons are either scalar or 
vector or iso-vector type. The quasi-particle interaction functions are 
related to the two-particle forward scattering via the exchange of any
one of these bosons. We further assume that the external magnetic field 
acts as a constant background and is present everywhere. In the next
subsection we consider the processes with scalar boson exchange. 

\subsection{Scalar Boson Exchange}
To compute the Landau-Fermi liquid interaction function $f$ from two particle 
forward scattering matrix with scalar boson ($\sigma$) exchange, we start 
with the definition of transition matrix element, given by 
\begin{eqnarray}
T_{fi}=i\int \rho_{s}(x)\phi(x)d^{4}x
\end{eqnarray}
where
\begin{eqnarray} 
\rho_{s}(x)=g_s\bar{\psi}(x)\psi(x)
\end{eqnarray}
is the scalar density and $\phi(x)$ is the scalar field satisfies the 
the well known Klein-Gordan equation
\begin{eqnarray}
(\partial^2+m^2)\phi(x)=\rho_s(x) 
\end{eqnarray}
The solution of this equation can be obtained by Green's function 
technique and is given by    
\begin{equation}
\phi(x)=-\int \frac{d^{4}q}{(2\pi)^4} 
\frac{\exp[-iq.(x-x^\prime)]}{q^2-\zeta^2}
\rho_{s}(x^\prime)d^4x^\prime
\end{equation}
where $q^\mu\equiv(q^0,\vec q)$ is the exchanged four momentum.

Now to evaluate $T_{fi}$, we substitute the Green's function solution
(eqn.14) into eqn.(12) to obtain $\rho_s(x)$ and finally putting it into
eqn.(11). We consider the following strong interaction processes:
\begin{eqnarray}
&(i)& ~~~p+p\longrightarrow p+p, \nonumber \\
&(ii)&~~~ n+n\longrightarrow n+n ~~{\rm and}~~\nonumber \\
&(iii)&~~~ p+n\longrightarrow p+n. \nonumber 
\end{eqnarray}
The basic diagrams are as follows:
\begin{figure} 
\psfig{figure=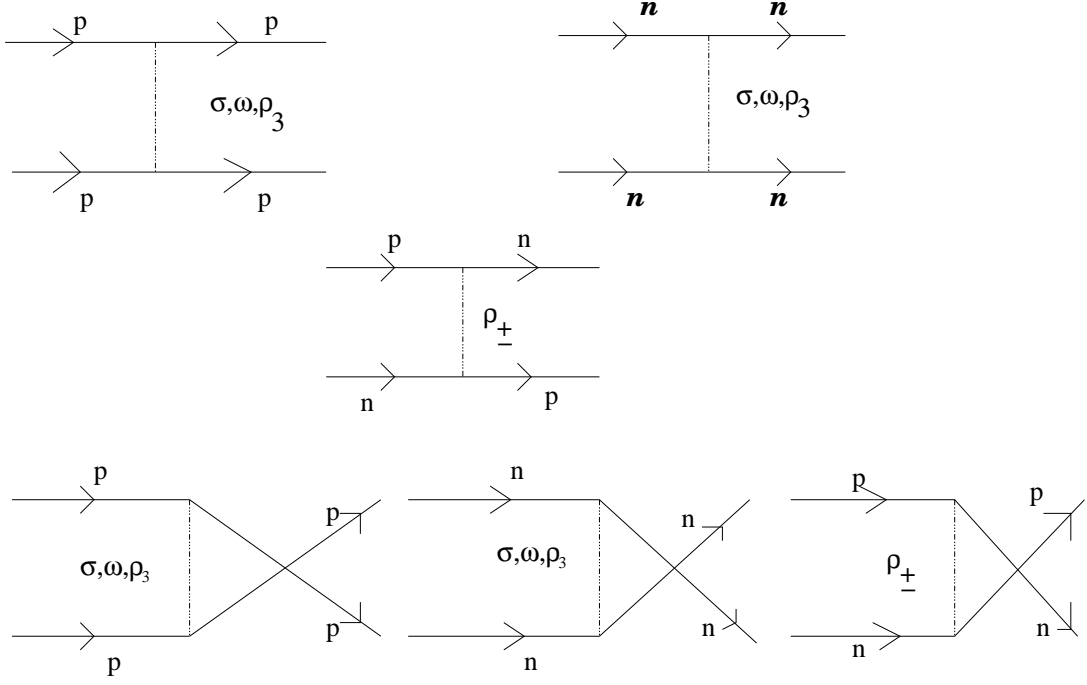,height=0.5\linewidth}
\caption{Basic Elementary Processes}
\end{figure}
We evaluate $T_{fi}$ for both direct and exchange type interactions for
all these processes. We first consider the scattering process represented by 
$(i)$. The transition matrix element is given by
\begin{equation}
T_{fi}=-i\int \rho_s(x) \frac{d^q}{(2\pi)^4}~
\frac{\exp[-iq(x-x^\prime)]}{q^2-\zeta^2} \rho_s(x^\prime)
d^4xd^4x^\prime
\end{equation}
Now using the up and down spin positive energy spinors for proton, we have
(since the temperature of the system $\ll$ the typical chemical potential for 
nucleons $\sim 1$GeV, we have not considered the negative energy states).
\begin{equation}
\rho_{s}(x^\prime)=\frac {g_s}{L_y
L_z}\exp[-i\{(E_{\nu_1}-E_{\nu_2})t^\prime-
(p_{1y}-k_{1y})y^\prime -(p_{1z}-k_{1z})z^\prime\}][\rho_s(x^\prime)]
\end{equation}
Then substituting for $\phi(x),~\rho_s(x)$  and $\rho_s(x^\prime)$ and 
integrating over $t^\prime,~y^\prime$ and $z^\prime$, we have the direct 
part for $p-p$ (represented by $(i)$) scattering matrix element.
\begin{eqnarray}
T_{fi,(pp\rightarrow pp)}^{(d,s)}&=&-i\int \frac{d^{4}q}{2\pi}
\delta(q^0-E_{\nu_1}+E_{\nu_2})
\delta(q_y-p_{1y}+k_{1y})\delta(q_z-p_{1z}+k_{1z})
\frac {\exp(iq.x)}{q^2-\zeta^2}\times \nonumber\\ &&[\rho(x^{'})]
\frac{g_s}{L_yL_z}\exp(iq_{x}x^\prime)dx^\prime\frac{g_s}{L_yL_z}
[\rho(x)]d^4x \nonumber \\ 
&&\exp[i\{(E_{\nu_1^\prime}-E_{\nu_2^\prime})t-(p_{2
y}-k_{2y})y-(p_{2z}-k_{2z})z\}]
\end{eqnarray}
Next integrating over $t,~y,~z$ and $q^0,~q_y,~q_z$ we have
\begin{eqnarray}
T_{fi,(pp\rightarrow pp)}^{(d,s)}&=&-i(2\pi)^3 \delta(E)\delta(p_y)
\delta(p_z)\frac{g_{s}^{2}}{(L_yL_z)^2}
\int \frac {dq_x}{2\pi}[\rho(x)][\rho(x^\prime)]
\exp[-iq_{x}(x-x^\prime)]\nonumber\\&&
\times\frac{1}{q^2-\zeta^2}dxdx^\prime
\end{eqnarray} 
where the $\delta$ functions represent the abbreviated form of energy and 
momentum conservations along $y$ and $z$ directions. Finally, evaluating the 
contour integral over $q_x$ and using the conditions for forward scattering 
given by
$E_{\nu_1}=E_{\nu_2}=E_{\nu}~~{\rm(say)}$;
$E_{\nu_1^\prime}=E_{\nu_{2}^\prime}=E_{\nu^\prime}~~{\rm(say)}$;
$p_{1y}=k_{1y}=p_y~~{\rm(say)}$; 
$p_{2y}=k_{2y}=p_y^\prime~~{\rm(say)}$; 
$p_{1z}=k_{1z}=p_z~~{\rm(say)~~ and}$
$p_{2z}=k_{2z}=p_z^\prime$ (say), we have the Landau-Fermi liquid interaction 
function corresponding to direct interaction
\begin{eqnarray}
f_{pp\rightarrow pp}^{(d,s)}=\frac{g_{s}^{2}}{8\zeta}\int 
\exp[-\zeta \left|x-x^\prime \right|][\rho(x)][\rho(x^\prime)]dx
dx^\prime
\end{eqnarray}
where
\begin{equation}
[\rho_s(x)]=\bar{u}(x,\nu_2,k_1)u(x,\nu_1,p_1)
\end{equation}
Then summing over the final spin states and averaging over the initial
spin states we have for direct diagram
\begin{equation}
[\rho_s(x)]=\frac{1}{2} {\rm{Tr}}[\Lambda(x,\nu_1,\nu_2,p_1,k_1)]
\end{equation}
and similarly for $x^\prime$. Whereas for the exchange diagram
\begin{equation}
[\rho_s(x)][\rho_s(x^\prime)]=\frac{1}{4} 
{\rm{Tr}}[\Lambda(x,\nu_1,\nu_2,p_1,k_1) 
\Lambda(x,\nu_1^\prime,\nu_2i^\prime,p_2,k_2)]
\end{equation}
where 
\begin{equation}
\Lambda(x,\nu_1,\nu_2,p_1,k_1)=\sum_{{\rm{spin}}}u(x,\nu_1,p_1)
\bar{u}(x,\nu_2,k_2)
\end{equation}

To obtain $\Lambda$'s for the process indicated by $(i)$, we consider the 
direct diagram for forward scattering. Then substituting the up and down spin 
solutions of Dirac equation for proton (eqns.(7) and (8)) in the expression 
for projection operator (eqn.(23)), we get
\begin{equation}
\Lambda=\frac{1}{2E_\nu}[Ak_\mu\gamma^\mu (\mu=0 ~~{\rm{and}}~~
z)\hfil\break+mA+B k_\mu\gamma^\mu (\mu=y ~~{\rm{and}}~~ p_y=p_\perp)]
\end{equation}
The matrices $A$ and $B$ are given by
\begin{equation}
A=\left ( \begin{array}{l c c r}I_\nu I_\nu^\prime &0&0&0 \\
0 & I_{\nu-1}I_{\nu-1}^\prime &0 &0 \\
0 & 0 &I_\nu I_\nu^\prime &0 \\
0 & 0 & 0 &I_{\nu-1}I_{\nu-1}^\prime \\
\end{array}
\right )
\end{equation}
\begin{equation}
B= \left ( \begin{array}{l c c r}I_{\nu-1} I_\nu^\prime &0&0&0 \\
0 & I_\nu I_{\nu-1}^\prime &0 &0 \\
0 & 0 &I_{\nu-1} I_\nu^\prime &0 \\
0 & 0 & 0 &I_\nu I_{\nu-1}^\prime \\
\end{array}
\right )
\end{equation}
where the primes indicate the functions of $x'$. To the best of our knowledge,
eqn.(24) is an entirely new result, first obtained in our previous
publication \cite{R25} and has not been reported earlier. Further, these 
results play key roles in all kinds of calculation related to the 
electromagnetic and weak interactions in presence of strong quantizing magnetic
field. In a recently completed work \cite{SCH} we have used this results to 
study chiral symmetry violation in presence of strong magnetic field.

Since $\gamma$ matrices are traceless and both $A$ and $B$ matrices are 
diagonal with identical blocks, it is very easy to evaluate the traces of 
the product of $\gamma$-matrices multiplied with any number of $A$ and/or $B$, 
from any side with any order e.g., 
\begin{equation}
{\rm{Tr}}(\gamma^\mu \gamma^\nu A_1A_2..B_1B_2..)={\rm{Tr}}(A_1A_2..B_1B_2..)
g^{\mu\nu},
\end{equation}
\begin{equation}
{\rm{Tr}}(\gamma^\mu\gamma^\nu\gamma^\lambda\gamma^\sigma
A_1A_2..B_1B_2..)={\rm{Tr}}(A_1A_2..B_1B_2..)(g^{\mu \nu}
g^{\sigma \lambda}-g^{\mu \lambda}g^{\nu \sigma}+g^{\mu \lambda}g^{\nu
\sigma}), 
\end{equation}
${\rm{Tr}}$(product of odd $\gamma$s with  $A$ and/or $B)=0$  etc. The other
interesting aspects of $A$ and $B$ matrices are:\\
i) $k_{1\mu}k^{2\mu}{\rm{Tr}}(A_1A_2)= (E_1E_2-k_{1z}k_{2z}){\rm{Tr}}(A_1A_2)$\\
ii) $k_{1\mu}k^{2\mu}{\rm{Tr}}(B_1B_2)= \vec k_{1\perp}.\vec
k_{2\perp}{\rm{Tr}}(B_1B_2)$\\
iii) $k_{1\mu}k^{2\mu}{\rm{Tr}}(A_1B_2)= k_{1\mu}k^{2\mu}{\rm{Tr}}(B_1A_2)=0$\\
iv) $p_{1\mu}k^{1\mu}p_{2\nu}k^{2\nu}{\rm{Tr}}(A_1B_2)\neq 0=
(E_{\nu_1}E_{\nu_2^\prime}-p_{1z}k_{1z})\vec p_{2\perp}.\vec k_{2\perp}
{\rm{Tr}}(A_1B_2)$

These set of relations are also totally new results and have been
derived in our previous publication \cite{R25}.

Then we have following Baym and Chin \cite{R26}
\begin{eqnarray}
f_{pp\rightarrow pp}^{(d,s)}=-\frac{g_{s}^{2}m^2}{16\zeta
E_{\nu}E_{\nu^\prime}}
\int \exp[-\zeta \left|x-x^\prime \right|]{\rm{Tr}}A {\rm{Tr}}A^\prime 
dx dx^\prime
\end{eqnarray}
where
\begin{eqnarray} 
{\rm{Tr}}A=2(I_{\nu}^{2}+I_{\nu-1}^{2})
~~{\rm and}~~
{\rm{Tr}}A^\prime=2(I_{\nu^\prime}^{2}+I_{\nu^\prime-1}^{2})
\end{eqnarray}

On the other hand if we consider the direct interaction corresponding to the
process $(ii)$, we have
\begin{eqnarray}
f_{nn\rightarrow nn}^{(d,s)}=-\frac{g_{s}^{2}m^2}{\zeta^2 EE^\prime}
\end{eqnarray}
This result is same as that obtained by Baym and Chin long ago.

Next we consider a bit complicated case; scattering process represented
by $(iii)$. In this case only protons are affected by strong quantizing
magnetic field, their Landau levels are populated. The transition matrix 
element is then given by
\begin{eqnarray}
T_{fi}=i\int \rho_{s}^{p}(x)\phi(x)d^{4}x
\end{eqnarray}
where $\rho_{s}^{(p)}(x)$ is the scalar density corresponding to proton 
component and is given by eqn.(20). The scalar field
\begin{eqnarray}
\phi(x)=-\int \frac{d^{4}q}{(2\pi)^{4}}\frac{exp[-iq.(x-x^\prime)]}
{q^2-\zeta^2}\rho_{s}^{(n)}(x^\prime)d^{4}x^\prime
\end{eqnarray}
where the scalar density corresponding to neutron component is given by
\begin{eqnarray} 
\rho_{s}^{(n)}(x)=g_s\bar \psi_n(x)\psi_n(x)=
\frac{g_s}{V}\exp[-i(p_2-k_2).x][\rho_s^{(n)}],
\end{eqnarray}
where
\begin{eqnarray}
[\rho_{s}^{(n)}]=\bar{u}(k_2)u(p_2),
\end{eqnarray}
here $u$ and $\bar{u}$ are the usual form of spinor and the adjoint for neutron,
given by
\begin{eqnarray}
u^{(s)}=\frac{1}{2E}\left( \begin{array}{c}
\chi^{(s)}\\ \frac{\vec{\sigma}.\vec{p}}{E+m}\chi^{(s)}
\end{array} \right)
\end{eqnarray}
where $s=1$ and $2$ corresponding to the up and down spin cases respectively, 
and are given by
\begin{equation}
\chi^{(1)}=\chi^{\uparrow}=\left( \begin{array}{c}
1\\0\
\end{array} \right)
\end{equation}
and
\begin{equation}
\chi^{(2)}=\chi^{\downarrow}=\left( \begin{array}{c}
0\\1\
\end{array} \right)
\end{equation}

Using all these expressions and integrating over $x$, we have
\begin{eqnarray}
T_{fi(p+n\rightarrow p+n)}^{(d,s)}&=&-i\int d^{4}q
\frac{\exp[iq.x^\prime]}{q^2-\zeta^{2}}\frac{g_s}{V}[\rho_{s}^{(n)}]
\frac{g_s}{L_yL_z}\exp[-i\{(E_{\nu_1}-E_{\nu_2})t^\prime-i(p_{1y}-k_{1y})
y^\prime- i(p_{1z}-k_{1z})z^\prime\}]\nonumber \\ &&[\rho_{s}^{(p)}]
\delta^{4}(q-p_2+k_2)d^{4}x^\prime
\end{eqnarray}
Further integrating over $t^\prime,~y^\prime,~z^\prime~~{\rm and}~~
q_x,~q_y,~q_z~,q_0$ finally we have for the forward scattering case
\begin{eqnarray}
f_{p+n\rightarrow p+n}^{(d,s)}=-\frac{g_{s}^{2}}{\zeta^{2}E_{\nu}E}m^2
\end{eqnarray}
where we have used the same expression as before for $[\rho_s^{(p)}(x)]$
whereas
\begin{equation}
[\rho_s^{(n)}]=\frac{1}{2E}(p_\mu\gamma^\mu+m)
\end{equation}
Here we have also used some obvious results as given below to obtain the
above expression (eqn.(40)).
\begin{equation}
(a)~{\rm{Tr}}({\not{p}} +m)=4m,~~(b)~{\rm{Tr}}(\Lambda)=m{\rm{Tr}}A^\prime
~~~{\rm{and}}~~~
(c)~\int_{-\infty}^{+\infty}{\rm{Tr}} A^\prime dx^\prime=4
\end{equation}
Although the structural forms of both eqns.(40) and (31) are identical, the 
appearance of proton energy in the denominator of eqn.(40) makes it 
qualitatively different from eqn.(31). This is also obvious from the physics 
point of view. The interaction with $\sigma$-meson exchange can not distinguish
neutron and proton. The difference comes from the change in proton parameters
(spinor wave functions, energy eigen values etc.) in presence of strong 
quantizing magnetic field. As we will see later that in the evaluation of 
energy densities for proton and neutron separately the form of momentum 
integrals (phase space integrals) will differ significantly (which has
also been discussed in the previous section).

We shall next evaluate the Landau interaction function for exchange part with 
$\sigma$-meson exchange. In this case the processes represented by $(i)$
and $(ii)$ give results in which there are nothing surprising. Whereas,
as we shall show that the Landau interaction function obtained from the
process $(iii)$ is complex in nature. Which will give complex self
energy of both proton and neutron. This is also true for exchange
diagrams with vector boson and neutral iso-vector boson exchange, whereas 
in the case of charged iso-vector exchange, the direct term gives
complex Landau interaction function and as a result the complex self-energies.
The findings are to the best of our knowledge are new and not been
reported earlier.

We start with the process $(i)$. In this case the forward scattering
condition gives
$E_{\nu_1}=E_{\nu_2^\prime}=E_\nu ~~{\rm{(say)}},~~ E_{\nu_1^\prime}=
E_{\nu_2}=E_{\nu^\prime} ~~{\rm{(say)}},~~ 
p_{1y}=k_{2y}=p_y ~~{\rm{(say)}},~~ p_{1z}=k_{2z}=p_z ~~{\rm{(say)}},~~ 
p_{2y}=k_{1y}=p_y^\prime ~~{\rm{(say)}},~~ p_{2z}=k_{1z}=p_z^\prime
~~{\rm{(say)}}$. Then we have the Landau interaction function
\begin{eqnarray}
f_{p+p\rightarrow
p+p}^{(ex,s)}&=&\frac{g_{s}^{2}}{32E_{\nu}E_{\nu^\prime}}\int dx
dx^\prime\frac{\exp[-k\left
|x-x^\prime\right|]}{K}[(E_{\nu}E_{\nu^\prime}-p_zp_{z^\prime}+m^2)
{\rm{Tr}}(AA^\prime)\nonumber
\\&&-\vec{p_{\perp}}.\vec{p_{\perp}^\prime}{\rm{Tr}}(BB^\prime)]
\end{eqnarray}
where $K=(q_{y}^2+q_{z}^2+\zeta^2)^{1/2},~q_{i}=(p_i-p_{i}^\prime)$ with
$i=y$ and $z$. The traces are obtained from eqns.(25) and (26).

Similarly for the process $(ii)$, the Landau quasi-particle interaction
function is given by
\begin{eqnarray}
f_{n+n\rightarrow n+n}^{(ex,s)}=\frac{g_{s}^{2}}{4EE^\prime}
\frac{p.p^\prime+m^2}{(p^\prime-p)^2-\zeta^2}
\end{eqnarray} 
and is identical with zero field result of Baym and Chin.

Next we consider the exchange term corresponding to the process $(iii)$
with scalar boson transfer. In this case the transition matrix element
is given by
\begin{equation}
T_{fi}=i\int \rho_s^{(pn)}(x)\phi(x) d^4x
\end{equation}
where 
\begin{eqnarray}
\rho_s^{(pn)}(x)&=&g_s\bar \psi_p(x)\psi_n(x)\nonumber \\
&=& \frac{1}{L_yL_z}\exp[-i\{(E_p-E_k)t-(p_y-k_y)y-(p_z-k_z)z\}]
\exp(ip_xx)[\rho_s^{(pn)}(x)]
\end{eqnarray}
and
\begin{equation}
\phi(x)=-\int \frac{d^4q}{(2\pi)^4}~\frac{\exp[-iq(x-x^\prime)]}{q^2
-\zeta^2} \rho_s^{(np)}(x^\prime d^4x^\prime
\end{equation}
Hence
\begin{eqnarray}
T_{fi}&=&-i\int \frac{d^q}{(2\pi)^4}~\frac{g_s}{L_yL_z}~\exp[-i\{
(E_{\nu_1}-E_{\nu_2^\prime})t -(p_{1_y}-k_{2_y})y- (p_{1_z}-k_{2_z})z\}]
[\rho_s^{(pn)}(x)] \nonumber \\ && \frac{\exp[-iq(x-x^\prime)]}{q^2-\zeta^2}~
 \frac{g_s}{L_yL_z} [\rho_s^{(np)}(x^\prime)]\nonumber \\
&&\exp[i\{
(E_{\nu_1^\prime}-E_{\nu_2})t^\prime -(p_{2_y}-k_{1_y})y- (p_{2_z}-k_{1_z})z\}]
\exp[-i(p_{1_x}-k_{1_x})x]
d^4xd^4x^\prime
\end{eqnarray}
Now integrating over $t$, $y$, $z$, 
$t^\prime$, $y^\prime$, $z^\prime$, $q_0$,
$q_x$, $q_y$ and $q_z$, we have the Landau quasi-particle interaction
function with the condition of forward scattering
\begin{equation}
f_{p+n\rightarrow p+n}^{(ex,s)}=\frac{g_{s}^{2}}{4} \int dx dx^\prime
\frac{\exp[i(q_x-p_x+p_x^\prime) (x-x^\prime)]}{(p-p^\prime)^2-\zeta^2}
{\rm{Tr}}[\Lambda_{pn}(x,p)\Lambda_{np}(x^\prime,p^\prime)]
\end{equation}
Now to evaluate the projection operators $\Lambda_{np}$ and
$\Lambda_{pn}$, we define
\begin{equation}
p_x+ip_y=p_\perp \exp(i\phi) ~~{\rm{and}}~~
p_x-ip_y=p_\perp \exp(-i\phi)
\end{equation}
Then from the ranges of $p_x~{\rm{and}}~ p_y$,
both of which vary from $-\infty$ to $+\infty$ and the range of
$p_\perp$, which is given by $(2\nu qB_m)^{1/2}$, with  $0 \leq \nu 
\leq\nu_{\rm{\max}}$ where $\nu_{\rm{max}}$ is a function of density of 
matter and the strength of magnetic field, we have $0 \leq \phi
\leq\pi$. With these choices, we have
\begin{eqnarray}
\Lambda_{np}(x,p)&=&u_n^\uparrow \bar u_p^\uparrow +
u_n^\downarrow \bar u_p^\downarrow\nonumber \\
&=& \frac{1}{2E}\left (\begin{array}{llccr}
I_\nu(E+m)& 0&-p_zI_\nu& -ip_\perp I_{\nu-1}\\
0 &I_{\nu-1}(E+m)& -ip_\perp I_\nu & p_zI_{\nu-1}\\ 
p_zI_\nu& p_\perp e^{-i\phi} I_{\nu-1} &
\frac{(ip_\perp^2e^{-i\phi}-p_z^2)}{E+m}I_\nu &
\frac{p_zp_\perp(e^{-i\phi}-i)}{E+m}I_{\nu-1}\\
p_\perp e^{i\phi} I_\nu & -p_zI_{\nu-1}&
-\frac{p_zp_\perp(e^{i\phi}+i)}{E+m}I_\nu&
-\frac{(p_z^2+ ip_\perp^2e^{i\phi}}{E+m}I_{\nu-1}
\end{array}
\right )
\end{eqnarray}
Which after some simplification reduces to
\begin{eqnarray}
\Lambda_{np}&=&\frac{1}{2E} \Big [(E\gamma_0+m-p_z\gamma_z){\cal{A}}
+\frac{p_\perp^2}{E+m} ~\frac{{\cal{A}}}{2} \Sigma_z(1-\gamma_0)-
\frac{p_zp_\perp}{2(E+m)} (1-\gamma_0)\Sigma_{yz}{\cal{A}}\nonumber \\
&+& \frac{iP_\perp^2}{E+m}~\frac{{\cal{A}}}{2}(1-\gamma_0)X
+\frac{p_zp_\perp}{E+m} \Sigma_{zx} X^* \frac{{\cal{A}}}{2} (1-\gamma_0)
\nonumber \\ &-& ip_\perp\frac{(1+\gamma_0)}{2}\gamma_x{\cal{A}}- p_\perp
\gamma_x \frac{{\cal{A}}}{2}(1+\gamma_0)X^*\Big ]
\end{eqnarray}
where
\begin{equation}
X=\left ( \begin{array}{lccr}
e^{-i\phi}& 0 & 0 &0 \\
0& e^{i\phi}& 0 & 0  \\
0& 0 &e^{-i\phi}&  0  \\
0& 0 & 0 &e^{i\phi}
\end{array}
\right ),
\end{equation}
\begin{equation}
\Sigma_{ij}=\left ( \begin{array}{lr}
\sigma_i\sigma_j & 0\\
0& \sigma_i \sigma_j 
\end{array}
\right ),
\end{equation}
\begin{equation}
{\cal{A}}=\left ( \begin{array}{l c c r}I_\nu &0&0&0 \\
0 & I_{\nu-1}&0 &0 \\
0 & 0 &I_\nu  &0 \\
0 & 0 & 0 &I_{\nu-1}
\end{array}
\right ),
\end{equation}
and
\begin{equation}
\Sigma_i=\left ( \begin{array}{lr}
\sigma_i & 0\\
0 & \sigma_i
\end{array}
\right )
\end{equation}
whereas, $\gamma$'s are Dirac matrices in standard representation and
$\sigma_i$ are the usual Pauli spin matrices. Similarly
\begin{eqnarray}
\Lambda_{pn}(x,p)&=&u_p^\uparrow \bar u_n^\uparrow +
u_p^\downarrow \bar u_n^\downarrow\nonumber \\
&=& \frac{1}{2E}\left (\begin{array}{lccr}
I_\nu(E+m)& 0&-p_zI_\nu& -p_\perp e^{-i\phi} I_{\nu}\\
0 &I_{\nu-1}(E+m)& -p_\perp e^{i\phi}I_{\nu-1} & p_zI_{\nu-1}\\ 
p_zI_\nu& ip_\perp I_{\nu} &
-\frac{[(E^2-m^2)-p_\perp^2(ie^{i\phi}-1)]}{E+m}I_\nu &
\frac{p_zp_\perp(i-e^{-i\phi})}{E+m}I_{\nu}\\
-ip_\perp I_{\nu-1} & -p_zI_{\nu-1}&
\frac{p_zp_\perp (i+e^{i\phi})}{E +m}I_{\nu-1}&
\frac{[p_\perp^2(ie^{-i\phi}+1)-(E^2-m^2)]}{E+m}I_{\nu-1}
\end{array}
\right )
\end{eqnarray}
After some simplification as before this can also be written as 
\begin{eqnarray}
\Lambda_{pn}&=&\frac{1}{2E} \Big [(E\gamma_0+m-p_z\gamma_z){\cal{A}}+
\frac{p_\perp^2}{2(E+m)}(1-\gamma_0)\Sigma_{z}{\cal{A}}+
\frac{ip_zp_\perp}{2(E+m)} (1-\gamma_0)\Sigma_{x}{\cal{A}}\nonumber \\
&+& \frac{ip_\perp^2}{E+m}\frac{{\cal{A}}}{2}(1-\gamma_0)X^*
+\frac{p_zp_\perp}{E+m} \Sigma_{yx} X \frac{{\cal{A}}}{2} (1-\gamma_0)
\nonumber \\ &+& p_\perp{{\cal{A}}}\Sigma_{x}X^*
-\frac{(1-\gamma_0)}{2}\Sigma{yz}{\cal{A}}\Big ]
\end{eqnarray}
Since diagonal elements for $\Sigma_i$, $\Sigma_{ij}$ and $\gamma_i$ with
$i\neq j$ and $i,j=x,y,z$, are zero,
the traces of all the quantities multiplied with these matrices are
identically zero. Therefore in both $\Lambda_{np}$ and
$\Lambda_{pn}$ only the first and the fourth terms will contribute in 
trace calculation. As a result the Landau quasi-particle interaction
function will become complex in nature. The explicit form of the trace
term is  given by
\begin{eqnarray}
&{\rm{Tr}}&[\Lambda_{np}(p^\prime)\Lambda_{pn}(p)]= \frac{1}{E_\nu
E_{\nu^\prime}} \{(E_\nu E_{\nu^\prime}+p_zp_z^\prime +m^2)2
(I_\nu I_{\nu^\prime}+I_{\nu-1}I_{\nu^\prime -1})
-\frac{p_\perp^2p_\perp^{\prime^2}}{4(E_\nu+m)(E_{\nu^\prime+m})}
\nonumber \\ && 4
(I_\nu I_{\nu^\prime}+I_{\nu-1}I_{\nu^\prime -1}) \nonumber \\
&-&\frac{iE_\nu p_\perp^{\prime^2}}{2(E_{\nu^\prime}+m)} 2
(I_\nu I_{\nu^\prime}\exp(i\phi)+I_{\nu-1}I_{\nu^\prime -1}\exp(-i\phi))
+\nonumber \\ &&\frac{i mp_\perp^{\prime^2}}{2(E_{\nu^\prime}+m)} 2
(I_\nu I_{\nu^\prime}\exp(i\phi)+I_{\nu-1}I_{\nu^\prime -1}\exp(-i\phi))
\nonumber \\
&-&\frac{iE_{\nu^\prime} p_\perp^2}{2(E_\nu+m)} 2
(I_\nu I_{\nu^\prime}\exp(-i\phi)+I_{\nu-1}I_{\nu^\prime -1}\exp(i\phi))
+\frac{im p_\perp^2}{2(E_\nu+m)} 2
(I_\nu I_{\nu^\prime}\exp(-i\phi)+\nonumber \\
&&I_{\nu-1}I_{\nu^\prime -1}\exp(i\phi)) ]
\end{eqnarray}
Obviously the last four  terms  in the above expression are  complex in 
nature.

Now the self-enegies for proton corresponding to the direct and exchange 
diagrams of process $(i)$ is given by
\begin{eqnarray}
\varepsilon_\nu^{(d/ex)}(p_z)&=&E_\nu(p_z)+\frac{1}{(2\pi)^2}
\sum_{\nu^\prime=0}^{[\nu_{\rm{max}}]}(2-\delta_{\nu^\prime 0})
 \int_{p_y^\prime=-\infty}^\infty
\int_{p_{z'}=-p_f}^{p_f} dp_y dp_{y'} dp_{z'}
f_{pp\rightarrow pp}^{(d/ex,s)} \nonumber \\
&=&E_\nu (p_z)+\Delta E_{\nu;{\rm{Dir./ex}}}^{(s)}(p_z)
\end{eqnarray}
For the process $(ii)$ the neutron self-energy corresponding to  direct or
exchange digram is given by
\begin{eqnarray}
\varepsilon^{(d/ex)}(p)&=&E(p)+\frac{g_n}{(2\pi)^3}\int
f_{n+n\rightarrow n+n}^{(d/ex,s)}d^3p^\prime\nonumber \\
&=&E(p)+\Delta E_{\rm{Dir./ex}}^{(s)}(p)
\end{eqnarray}
Next for the process $(iii)$ with direct diagram, the proton self-energy 
is given by
\begin{eqnarray}
\varepsilon_\nu^{(d)}(p_z)&=&E_\nu(p_z)+\frac{eB_m}{2\pi^2}
\sum_{\nu^\prime=0}^{[\nu_{\rm{max}}]}(2-\delta_{\nu^\prime 0})\int
f_{p+n\rightarrow p+n}^{(d,s)}dp_z^\prime \nonumber \\
&=&E_\nu(p_z)+\Delta E_{\nu;\rm{Dir./ex}}^{(s)}(p_z)
\end{eqnarray}
Similarly for neutron, it is given by
\begin{eqnarray}
\varepsilon^{(d)}(p)&=&E(p)+\frac{g_n}{(2\pi)^3}\int
f_{p+n\rightarrow p+n}^{(d,s)}d^3p^\prime\nonumber \\
&=&E(p)+\Delta E_{\rm{Dir.}}^{(s)}(p)
\end{eqnarray}
Finally the exchange part of self-energy corresponding to process
$(iii)$ are same for both proton and neutron and it is given by
\begin{eqnarray}
\varepsilon_\nu^{(ex)}(p_z)&=&E_\nu(p_z)+
\sum_{\nu^\prime=0}^{[\nu_{\rm{max}}]}(2-\delta_{\nu^\prime 0}) \int_0^\pi
d\phi d\phi^\prime p_\perp p_\perp^\prime \cos(\phi)\cos(\phi^\prime)
\int_{-p_F}^{+p_F}dp_z^\prime
f_{p+n\rightarrow p+n}^{(ex,s)}\nonumber \\
&=&E_\nu(p_z)+\Delta E_{\nu;\rm{ex}}^{(s)}(p_z)
\end{eqnarray}
This last expression for self energy for both proton and neutron is complex 
in nature. As we will see in the subsequent sections that similar
results can also be obtained with vector and iso-vector exchange
type interactions.
\subsection{Vector Boson Exchange}
We shall now consider the vector boson exchange type interaction.
The transition matrix element is given by
\begin{eqnarray}
T_{fi}=i\int j^{\mu}(x)V_{\mu}(x)d^{4}x 
\end{eqnarray}
where
\begin{eqnarray}
j^{\mu}(x)=g_{v}\bar{\psi}(x)\gamma^{\mu}\psi(x)
\end{eqnarray}
is the particle current and the vector $V^{\mu}(x)$ satisfies the 
Klein-Gordan equation given by
\begin{eqnarray}
(\partial^2+\eta^2)V^{\mu}(x)=j^{\mu}(x)
\end{eqnarray}
Hence the Green's function solution for $V^{\mu}(x)$ is given by
\begin{eqnarray}
V^{\mu}(x)=-\int \frac{d^4q}{(2\pi)^4}\frac{\exp[-iq.(x-x^\prime)]}
{q^2-\eta^2}j^{\mu}(x^\prime)d^4x^\prime
\end{eqnarray}
where $\eta$ is the vector boson mass.

Let us first consider the direct term of the process $(i)$. Using the spinor 
solutions for proton (eqns.(7) and (8)), we have
\begin{eqnarray}
j^{\mu}(x)=\frac{g_s}{L_yL_z}\exp[-i\{(E_{\nu_1}-E_{\nu_2})t-(p_{1y}-k_{1y})y-
(p_{1z}-k_{1z})z\}][j^{\mu}]
\end{eqnarray}
where in the direct term we define
\begin{eqnarray}
[j^{\mu}]&=&\frac{1}{2}\sum_{{\rm{spin}}}\bar{u}(x^\prime,\nu_2,k_1)\gamma_\mu
u(x^\prime,\nu_1,p_1)
\nonumber\\&=&
\frac{1}{2}{\rm{Tr}}[\Lambda(x^\prime,\nu_1,\nu_2,p_1,k_1)\gamma^{\mu}]
\end{eqnarray}
whereas in the case of exchange term we use
\begin{equation}
[j^\mu(x)][j_\mu(x^\prime)]=\frac{1}{4}{\rm{Tr}}[\Lambda(x,\nu_1,\nu_2,p_1,k_1)
\gamma^\mu
\Lambda(x^\prime,\nu_1^\prime,\nu_2^\prime,p_2,k_2)
\gamma_\mu]
\end{equation}
Similar to scalar meson exchange case as discussed before, we first integrate
over $t^\prime,~y^\prime,~z^\prime$; then over $q_0,~q_y,~q_z$ and the 
contour integral is evaluated over $q_x$. The final integrations are over 
$t,~y$ and $z$. Thus the Landau interaction function for $p-p$ (process
$(i)$) forward scattering with vector boson exchange direct term is given by
\begin{eqnarray}
f_{p+p\rightarrow p+p}^{(d,v)}&=&\frac{g_{v}^2}{32\eta
E_{\nu}E_{\nu^\prime}}
\int\exp[-\eta\left |x-x^\prime\right|] dx
dx^\prime[(E_{\nu}E_{\nu^\prime}-p_zp_{z^{'}}){\rm{Tr}}A{\rm{Tr}}A^\prime
\nonumber\\ &&
+\vec{p_{\perp}}.\vec{p_{\perp}^\prime}{\rm{Tr}}B{\rm{TrB}}^\prime]
\end{eqnarray}
The above expression for process $(ii)$ with vector boson exchange
direct interaction is identical with the zero field case and is given by
\begin{eqnarray}
f_{n+n\rightarrow
n+n}^{(d,v)}=\frac{g_{v}^{2}}{\eta^2}\frac{1}{EE^\prime}(EE^\prime-
\bar {p}.\bar{p^\prime})
\end{eqnarray}
Next, for the process represented by $(iii)$, the matrix element is
given by
\begin{eqnarray}
T_{fi}=i\int j^{\mu}_{(n)}(x)V_{\mu}(x)d^{4}x 
\end{eqnarray} 
In this case the vector field is given by
\begin{eqnarray}
V^{\mu}(x)=-\int \frac{d^4q}{(2\pi)^4}\frac{\exp[-iq.(x-x^\prime)]}
{q^2-\eta^2}j^{(p)}_{\mu}(x^\prime)d^4x^\prime
\end{eqnarray}
where
\begin{eqnarray}
j_{(n)}^{\mu}(x)&=& g_v \bar \psi_n(x)\gamma^\mu \psi_n(x) \nonumber \\ &=&
\frac{g_v}{V}\exp[-i(p_2-k_2).x][j_{(n)}^{\mu}]
\end{eqnarray}
and
\begin{eqnarray}
j_{\mu}^{(p)}(x^\prime)=\frac{g_v}{L_yL_z}\exp[-i\{(E_{\nu_1}-
E_{\nu_2})t^\prime-(p_{1y}-k_{1y})y^\prime-(p_{1z}-k_{1z})z^\prime\}]
[j_{\mu}^{(p)}]
\end{eqnarray}
Further for the direct term
\begin{eqnarray} [j_{(n)}^{\mu}]=\frac{1}{2}{\rm{Tr}}[\bar{u}(k_2)\gamma^{\mu}u(p_2)]
\end{eqnarray}
and
\begin{eqnarray}
[j_{\mu}^{(p)}]=\frac{1}{2}{\rm{Tr}}[\bar{u}(k_1)\gamma_\mu u(p_1)]
\end{eqnarray}
Substituting all these terms in $T_{fi}$ and subsequently integrating over 
$x^\prime,~ t^\prime,~y^\prime,~z^\prime,$ $ x,~ t,~ y,~ z 
~~{\rm{and}}~~d^4q$, we have the direct 
type Landau-Fermi interaction function for $\omega$ meson exchange as 
\begin{eqnarray}
f_{p+n\rightarrow p+n}^{(d,v)}=-\frac{g_{v}^{2}}{4E_{\nu}E_{\nu^\prime}
\eta^2}p.p^\prime\int_{-\infty}^{+\infty} dx^\prime {\rm{Tr}}(A+B)
\end{eqnarray} 
where we have used
\begin{eqnarray}
[j_{(n)}^{\mu}(x)]=\frac{1}{4E}{\rm{Tr}}[({\not{p}}+m)\gamma^{\mu}]=
\frac{p^{\mu}}{E} 
\end{eqnarray}
and
\begin{eqnarray} 
[j_{\mu}^{(p)}(x^\prime)]=\frac{1}{4E}{\rm{Tr}}[(Ap^{\lambda}\gamma_{\lambda}+
Am+Bp^{\lambda}\gamma_{\lambda})\gamma_{\mu}]=\frac{1}{4E}{\rm{Tr}}[A+B]p_{\mu}
\end{eqnarray}
It can now be shown easily that 
\begin{eqnarray} 
\int_{-\infty}^{+\infty} dx^\prime {\rm{Tr}}
A(x^\prime)=4,~~\int_{-\infty}^{+\infty} dx^\prime {rm{Tr}}B(x^\prime)=0
\end{eqnarray}
Therefore
\begin{eqnarray}
f_{p+n\rightarrow p+n}^{(d,v)}=-\frac{g_{v}^{2}}{E_{\nu}E^\prime
\eta^2}(E^\prime E_{\nu}-p_{z}^\prime p_{z})
\end{eqnarray}
The structural form is same as that of non-magnetic case but qualitatively 
the result is entirely different as we have already discussed.

Now we shall consider exchange interaction terms corresponding to 
$\omega$-meson transfer. For the process $(i)$, i.e., $p+p\rightarrow p+p$, 
we have 
\begin{eqnarray}
f_{p+p\rightarrow
p+p}^{(ex,v)}&=&-\frac{g_{v}^{2}}{16E_{\nu}E_{\nu^\prime}}
\int_{-\infty}^{+\infty}\int_{-\infty}^{+\infty}  dx dx^\prime
\frac{\exp[-k\left |x-x^\prime\right|]}{K}\times \nonumber \\ 
&&[(p_zp_{z^\prime}-E_\nu E_{\nu^\prime}+2m^2){\rm{Tr}}(AA^\prime)-
2\vec p_\perp.\vec p_\perp^\prime {\rm{Tr}}(BB^\prime)]
\end{eqnarray}
In the case of  $n+n\rightarrow n+n$, i.e., process $(ii)$ the result is 
same as that of non-magnetic case and is given by
\begin{eqnarray}
f_{n+n\rightarrow
n+n}^{(ex,v)}=-\frac{g_{v}^{2}}{2}\frac{2m^2-p.p^\prime}{EE^\prime}
\frac{1}{(p-p^\prime)^2-\eta^2}
\end{eqnarray}
This is exactly the same as that of zero field result.

Let us now consider the exchange term corresponding to process $(iii)$.
In this case the calculation is almost identical with the scalar
exchange model. Here unlike the scalar case we have to evaluate 
${\rm{Tr}}[\Lambda_{np}(p)\gamma^\mu\Lambda_{pn}(p^\prime)\gamma_\mu]$. 
where the projection operators are given by eqns.(52) and (58).
We have noticed that from the non-zero contributions it is very easy to show 
\begin{eqnarray}
&&{\rm{Tr}}[\Lambda_{pn}(p)\gamma^0\Lambda_{np}(p^\prime)\gamma_0]= 
-{\rm{Tr}}[\Lambda_{pn}(p)\gamma^x\Lambda_{np}(p^\prime)\gamma_x]= 
-{\rm{Tr}}[\Lambda_{pn}(p)\gamma^y\Lambda_{np}(p^\prime)\gamma_y]\nonumber 
\\ &=& 
-{\rm{Tr}}[\Lambda_{pn}(p)\gamma^z\Lambda_{np}(p^\prime)\gamma_z]= \nonumber \\
&=&\frac{1}{4E_\nu E_{\nu^\prime}}[(E_\nu E_{\nu^\prime}+m^2+p_zp_z^\prime)
+\frac{ip_\perp^2}{2(E_\nu+m)}(m-E_{\nu^\prime})
{\rm{Tr}}[{\cal{A}}X{\cal{A}}^\prime]
+\frac{ip_\perp^{2^\prime}}{2(E_{\nu^\prime}+m)}(m-E_\nu){\rm{Tr}}[{\cal{A}}
{\cal{A}}^\prime
X^\prime]\nonumber \\
&-&\frac{p_\perp^2p_\perp^{\prime^2}}{2(E_\nu+m)(E_{\nu^\prime}+m)}
{\rm{Tr}}[{\cal{A}}{\cal{A}}^\prime]
\end{eqnarray}
where 
\begin{eqnarray}
&& {\rm{Tr}}[{\cal{A}}{\cal{A}}^\prime]=2(I_\nu I_{\nu^\prime} -I_{\nu-1}
I_{\nu^\prime-1})\nonumber \\
&& {\rm{Tr}}[{\cal{A}}{\cal{A}}^\prime X]=2(I_\nu I_{\nu^\prime}\exp(-i\phi) -I_{\nu-1}
I_{\nu^\prime-1}\exp(i\phi))\nonumber \\
&& {\rm{Tr}}[{\cal{A}}{\cal{A}}^\prime X]=2(I_\nu I_{\nu^\prime}\exp(i\phi^\prime) -I_{\nu-1}
I_{\nu^\prime-1}\exp(-i\phi^\prime))
\end{eqnarray}
It is quite obvious that complex contribution is also coming to the Landau 
interaction function and as a consequence to the quasi-particle energy
if we consider the exchange diagram with vector boson transfer in the process 
represented by $(iii)$ 

\subsection{Iso-Vector Boson Exchange}
Now, we consider the last part of our investigation; the iso-vector
boson exchange in neutron star matter. Since this part was not included in 
the original work of Baym and Chin on conventional relativistic version of 
Landau theory of Fermi liquid, we shall first develop a formalism for the 
non-magnetic case. Here, as we have mentioned, the iso-vector field 
$\vec{\rho^\mu}$ behaves like a four vector (indicated by symbol $\mu$)
in configuration space and a three vector (indicated by $\rightarrow$) in 
iso-spin space.

We start with the simplest case, the exchange of neutral iso-vector boson
$\rho^{\mu}_{3}$. The corresponding transition matrix element is given by
\begin{eqnarray}
T_{fi}&=&i\int j_{3}^{\mu}(x)\rho_{3,\mu}(x)d^{4}x \\
&&{\rm where}\nonumber \\
\rho_{3,\mu}(x)&=&-\int \frac{d^4q}{(2\pi)^4}
\frac{\exp[-iq.(x-x^\prime)]}{q^2-\theta^2}j_{3,\mu}(x^\prime)d^4x^\prime
\end{eqnarray}
where $\theta$ is the mass of iso-vector boson and 
\begin{eqnarray}
j^{\mu}_{3}(x)=g_{\rho}\bar{\psi}(x)\gamma^{\mu}\frac{\tau_3}{2}\psi(x)
\end{eqnarray} 
is the third component of iso-vector fermion current. In the case of neutral 
$\rho$-meson exchange
\begin{eqnarray}
j^{\mu}_{3}(x)=\frac{1}{2}[j^{\mu}_{3,(p)}(x)-j^{\mu}_{3,(n)}(x)]
\end{eqnarray}
where symbols $p$ and $n$ represent proton and neutron components respectively.
Using the conventional spinor solutions for both proton and neutron we
have,
\begin{eqnarray}
j^{\mu}_{3}(x)=\frac{1}{2}[\bar{u_{p}}(k)\gamma^{\mu}u_{p}(p)-
\bar{u_{n}}(k)\gamma^{\mu}u_{n}(p)]\exp[-i(p-k).x]
\end{eqnarray}
On substituting $j_{3}^{\mu}~~{\rm and}~~\rho_{3}^{\mu}$ in the transition 
matrix element and following the procedure by Baym and Chin for the 
case of vector meson exchange, we have the direct type Landau
quasi-particle interaction function with neutral iso-vector boson
exchange
\begin{eqnarray}
f^{(d,iv)}=-\frac{g_\rho^2}{4\theta^2}\frac{p.p^\prime}
{4EE^\prime}
\end{eqnarray}
Similarly, for exchange type interaction  we have 
\begin{eqnarray}
f^{(ex,iv)}=-\frac{g_\rho^2}{8EE^\prime}\frac{[2m^2-p.p^\prime]}
{[(p-p^\prime)^2-\theta^2]}
\end{eqnarray}
Now in the non-magnetic case, it is very easy to show that the same expressions 
as above (eqns.(94) and (95)) will also be obtained for the charged iso-vector
mesons exchange for the valid process represented by $(iii)$.

Let us now consider the presence of strong quantizing magnetic field, the 
Landau interaction functions for neutral $\rho$-meson exchange are almost
identical with the results of vector boson exchange (except a factor of
$1/2$). For the process $(i)$, with direct type interaction, we have
\begin{eqnarray}
f_{p+p\rightarrow p+p}^{(d,iv)}&=&\frac{g_{v}^2}{64\theta
E_{\nu}E_{\nu^\prime}}
\int\exp[-\theta\left |x-x^\prime\right|] dx
dx^\prime[(E_{\nu}E_{\nu^\prime}-p_zp_{z^{'}}){\rm{Tr}}A{\rm{Tr}}A^\prime
\nonumber\\ &&
+\vec{p_{\perp}}.\vec{p_{\perp}^\prime}{\rm{Tr}}B{\rm{TrB}}^\prime]
\end{eqnarray}
For the process $(ii)$, the direct interaction is again $1/2$ times the
non-magnetic result which we have already discussed
and in the case of $(iii)$ the Landau quasi-particle interaction
function will again be $1/2$ of that of  vector boson case. 
In the last case, we again got complex contribution in Landau
interaction function.

Next we consider the exchange of charged $\rho$-meson. For the sake of
convenience we define
\begin{eqnarray}
\rho_1^\mu(x)=\frac{1}{2}[\rho_+^\mu(x)+\rho_-^\mu(x)]~~{\rm and}~~
\rho_2^\mu(x)=\frac{1}{2}[\rho_+^\mu(x)-\rho_-^\mu(x)]
\end{eqnarray}
where $\rho_\pm^\mu(x)$ are related to the well known Ladder operators
in iso-spin space (causes $p\leftrightarrow n$). Then in the lowest order 
diagram, the only valid scattering process is as represented by $(iii)$. The 
transition matrix element for non-magnetic case is given by
\begin{eqnarray}
T_{fi}=-\frac{i}{16}\int \frac{d^4q}{(2\pi)^4}
&&(j^{(pn)}_{\mu}(x)+j^{(np)}_{\mu}(x))\frac{\exp[-iq.(x-x^\prime)]}
{q^2-\theta^2}\nonumber \\ && (j_{(pn)}^{\mu}(x)
+j_{(np)}^{\mu}(x))d^{4}xd^{4}x^\prime
\end{eqnarray} 
where 
\begin{eqnarray}  
j_{(pn)}^\mu(x)=g_\rho\bar{\psi_n}(x)\gamma^\mu\psi_p(x)
\end{eqnarray} 
and
\begin{eqnarray}  
j_{(np)}^\mu(x)=g_\rho\bar{\psi_p}(x)\gamma^\mu\psi_n(x)
\end{eqnarray}
After some simple algebraic steps we have
\begin{eqnarray}
T_{fi}&=&-\frac{i}{16}\int \frac{d^4q}{(2\pi)^4}[j^{(pn)}_{\mu}(x)
\frac{\exp[-iq.(x-x^{'})]}{q^2-\theta^2}j_{(np)}^{\mu}(x^{'})\nonumber\\
&+&j^{(np)}_{\mu}(x)\frac{\exp[-iq.(x-x^{'})]}{q^2-\theta^2}
j_{(pn)}^{\mu}(x^{'})]d^{4}xd^{4}x^{'} 
\end{eqnarray} 
Using the conventional form of spinor solutions for both proton and neutron, 
we have 
\begin{eqnarray}
j_{(pn)}^\mu(x)=\frac{g_\rho}{V}\exp[-i(p_p-p_n).x]
\bar{u_n}(p_n)\gamma^\mu u_p(p_p)
\end{eqnarray} 
and
\begin{eqnarray}
j^{(np)}_\mu(x^\prime)=\frac{g_\rho}{V}\exp[-i(p_n^\prime-p_p^\prime).x^{'}]
\bar{u_p}(p_p^\prime)\gamma_\mu u_n(p_n^\prime)
\end{eqnarray}
Then integrating over $d^4x,~~d^4x^\prime ~~{\rm and }~~d^4q$,
the Landau-Fermi interaction function for direct type forward scattering
\begin{eqnarray}
f^{(d,iv)}=-\frac{g_\rho^{2}p.p^\prime}{8EE^\prime\theta^2}
\end{eqnarray}
Similarly for the exchange case
\begin{eqnarray}
f^{(ex,iv)}=\frac{g_\rho^2(4m^2+p.p^\prime)}{16EE^\prime [(p-p^\prime)^2-
\theta^2]}
\end{eqnarray} 

Let us now consider the effect of strong quantizing magnetic field on
this process with charged $\rho$-meson exchange. In this case not only
the Dirac spinor solutions of proton, but the charged meson propagators will 
also get modified. The meson propagator is given by
\begin{eqnarray}
G_{\rho_{\pm}}(x)&=&\frac{eB_m}{4\pi^2}\sum_\nu \int dE_0\Phi^*(x^\prime)\Phi(x)
\exp[-i\{E_0(t-t^\prime)-
q_y(y-y^\prime) -q_z(z-z^\prime)\}]\nonumber \\ &&
\frac{1}{E_0-(\epsilon_q-i\delta)}
\frac{1}{4}[j_{pn}(x^\prime)-j_{np}(x^\prime)] d^4x^\prime
\end{eqnarray} 
where 
\begin{eqnarray}
\Phi(x)=I_\nu(x,p)=\left (\frac{qB_m}{\pi}\right )^{1/4}\frac{1}{(\nu !)^{1/2}}
2^{-\nu/2}
\exp \left [{-\frac{1}{2}qB_m\left (x-\frac{p_y}{qB_m} \right )^2}\right 
]\nonumber \\
H_\nu \left [(qB_m)^{1/2}\left (x-\frac{p_y}{qB_m} \right) \right ],
\end{eqnarray}
Then it is obvious that $\Phi^*(x)=\Phi(x)$. The quantity $\epsilon_q=
(q_z^2+\theta^2+2\nu eB_m)^{1/2}$ is the modified form of charged 
$\rho$-meson energy eigen value, obtained from the solution of Klein-Gordon 
equation in presence of strong quantizing magnetic field,
with $2\nu=2n+1+2\alpha$, $\alpha=-1,0$ or $+1$ and $n=0,1,2,..$, the
Landau principal quantum number.

Now in the direct case, unlike the scalar and vector mesons exchange, we
have to evaluate the product ${\rm{Tr}}[\Lambda_{pn}\gamma^\mu]
{\rm{Tr}}[\Lambda_{np}\gamma_\mu]$. As before, we calculate  this
quantity term by term (i.e., for $\mu=0,x,y$ and $z$).
Using the expressions for the projection operators, we have
\begin{eqnarray}
{\rm{Tr}}[\Lambda_{pn} \gamma^0]{\rm{Tr}}[\Lambda_{np}\gamma_0]
=\left [\frac{1}{2}{\rm{Tr}}{\cal{A}}-
\frac{ip_\perp^2}{4E_\nu(E_\nu+1)}{\rm{Tr}}({\cal{A}}X)\right ] 
\left [\frac{1}{2}{\rm{Tr}}{\cal{A}}^\prime-
\frac{ip_\perp^{\prime^2}}{4E_\nu^\prime(E_\nu^\prime+1)}{\rm{Tr}}
({\cal{A}}^\prime X^*)\right ] 
\end{eqnarray}
\begin{eqnarray}
{\rm{Tr}}[\Lambda_{pn} \gamma^x]{\rm{Tr}}[\Lambda_{np}\gamma_x]
={\rm{Tr}}[\Lambda_{pn} \gamma^y]{\rm{Tr}}[\Lambda_{np}\gamma_y]
=0
\end{eqnarray}
and
\begin{eqnarray}
{\rm{Tr}}[\Lambda_{pn} \gamma^z]{\rm{Tr}}[\Lambda_{np}\gamma_z]
=\left [-\frac{1}{2}{p_z\rm{Tr}}{\cal{A}} \right ] 
\left [-\frac{1}{2}{p_z^\prime\rm{Tr}}{\cal{A}}^\prime \right ] 
\end{eqnarray}
Obviously, the term with $\mu=0$ is complex in nature and it will make the
Landau quasi-particle interaction function and as a consequence the
quasi-particle energies for both neutron and proton complex in nature.

Fortunately, in the case of exchange type interaction with charged
iso-vector meson exchange, we have to evaluate
${\rm{Tr}}[\Lambda_{pp}\gamma^\mu \Lambda_{nn}\gamma_\mu]$ which gives
real contribution to quasi-particle energy.

To obtain the variation of self energies (both real and imaginary parts) we
have evaluated numerically the total self energy density for nucleons (both
protons and neutrons separately) then plotted the average proton self energy
as a function of proton density and are shown in the following two figures.
To draw these figures, we have solved a few non-linear equations
self-consistently. We assume that before the interaction is switched on, the
matter is in $\beta$-equilibrium, given by
\[ 
\mu_n=\mu_p+\mu_e
\]
(Which is obtained from the weak processes $n\longrightarrow p+e^-+\bar\nu_e$ 
and $p+e^-\longrightarrow n+\nu_e$, known as the URCA and the modified URCA
processes).
We further assume the charge neutrality condition, given by
\[
n_p=n_e
\]
and assuming that the baryon number density $n_B=n_p+n_n$ remains constant.
To solve for the chemical potentials $\mu_i$, where $i=n,p$ and $e$, we use
\[
n_n=\frac{(\mu_n^2-m_n^2)^{3/2}}{6\pi^2},
\]
\[
n_p=\frac{eB_m}{2\pi^2}\sum_{\nu=0}^{[\nu_{\rm{max}}]}
(2-\delta_{\nu 0})(\mu_p^2-m_p^2-2\nu eB_m)^{1/2}
\]
and
\[
n_e=\frac{eB_m}{2\pi^2}\sum_{\nu=0}^{[\nu_{\rm{max}}]}
(2-\delta_{\nu 0})(\mu_e^2-m_e^2-2\nu eB_m)^{1/2}
\]
Hence, knowing the chemical potentials, we obtained numerically the 
total free energies (both real and imaginary parts) per particle for both protons 
and neutrons.

\begin{figure} 
\psfig{figure=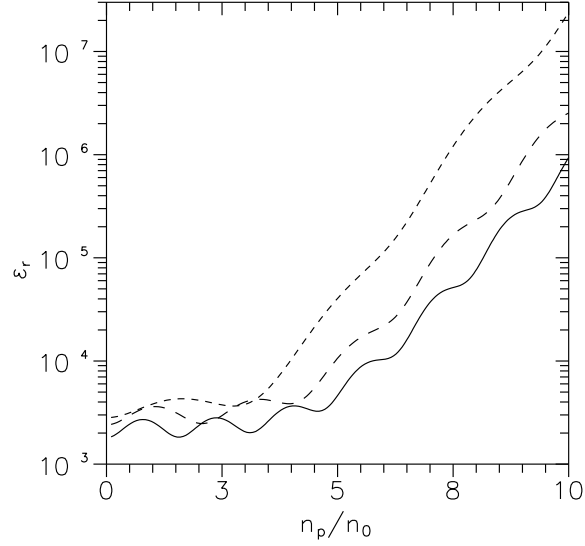,height=0.4\linewidth}
\caption{Variation of the real part of average proton self energy (in MeV.)
with proton
density (expressed in terms of normal nuclear density $n_0=0.17$fm$^{-3}$).
The upper curve is for $B_m=10^{14}$G, middle one is for $B_m=10^{16}$G and
the lower one is for $B_m=10^{18}$G.}
\end{figure}

\newpage
\begin{figure} 
\psfig{figure=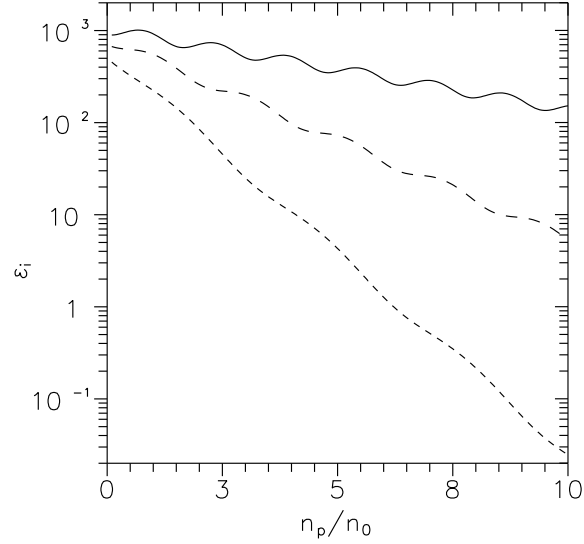,height=0.4\linewidth}
\caption{Variation of the imaginary part of average proton and neutron self 
energies (in MeV.)
with proton density (expressed in terms of normal nuclear density 
$n_0=0.17$fm$^{-3}$). The upper curve is for $B_m=10^{18}$G, middle one is for 
$B_m=10^{16}$G and the lower one is for $B_m=10^{14}$G.}
\end{figure}
As one can see from these two figures that the real part always increases
with the increase of matter density but the matter becomes more stable if the
magnetic field is stronger. Which is also true in all other cases, e.g.,
atoms, quark matter etc. \cite{Li,R10,Sc} in strong quantizing magnetic fields. 
Whereas the imaginary part decreases with the increase of matter density and at 
very high density the imaginary part is almost negligibly small with respect to
the corresponding real part. The reason is that at very high density, the
effect of Landau levels become insignificant, in other ward, at high enough
density the magnetic field no longer has quantum mechanical effect. On the
other hand at low density, it is significant and the effect is more as the
magnetic field becomes stronger. 

For the sake of completeness we have also plotted the life time of nucleons
(both neutron and proton) as a function of proton density, for three
different magnetic field strengths. The life times are in the strong
interaction time scale within a broad density range. Because of such small
time scale we belive that the bulk neutron star matter remains stable against strong
decay, just like a nucleus is a stable object although strong interaction
processes are going on inside. The other extreme could be that the magnetic
field strength at the core region of magnetars are much smaller than the
quantum critical value and the observed strong magnetic fields are concentrated
at the crustal region.

\begin{figure} 
\psfig{figure=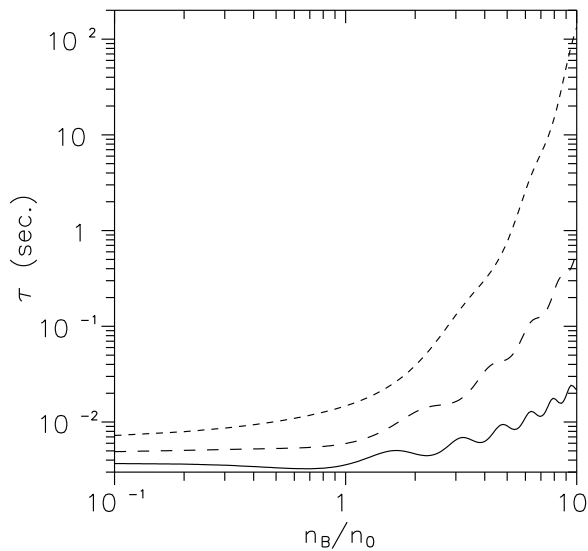,height=0.4\linewidth}
\caption{Variation of proton and neutron life times (in sec. in unit of
$10^{22}$)  with proton
density (expressed in terms of normal nuclear density $n_0=0.17$fm$^{-3}$).
The upper curve is for $B_m=10^{18}$G, middle one is for $B_m=10^{16}$G and
the lower one is for $B_m=10^{14}$G.}
\end{figure}
\section{Conclusions}
We therefore conclude that in the relativistic version of Landau theory of Fermi
liquid in presence of strong quantizing magnetic field, although the 
self-energies for both neutron and proton become complex in nature, the bulk 
neutron star matter still remain stable against strong decay. We have noticed from 
the complex nature of self-energies for both the dressed protons and neutrons, 
that the life time is within the strong interaction scale and we belive that 
just like a stable nucleus, the bulk neutron star matter will also remain stable in
presence of strong quantizing magnetic field.

The other explanation, which is of course not so convincing is that, we can have a 
stable dense neutron matter at the central
region of a magnetar with a  very low magnetic field strength.
The observed strong magnetic field in such objects are concentrated 
at the crustal region. However, in our opinion, the best explanation to have a 
magnetic field of strength $\sim 10^{14}$G at the surface and several orders of
magnitude large field at the core is that the conducting (and possibly not
superconducting because the field strength is $>$ the critical value) neutron star 
matter does not allow the magnetic flux lines to move from the core region to the 
crust. Since the density of neutron star matter at the core region is high
enough, a large number of Landau levels will be populated, even if the field
strength is extremely high. Which further means, that the quantum mechanical
effect of strong magnetic field will be extremely weak.

Finally, we would like to add a few words to compare the results, in
particular the complex nature of nucleon self energies we obtained in this
work with some of the well known phenomena already discussed in the
literature. 

About sixty years ago L.D. Landau had shown the possibility of some
kind of damping or dissipation in collisionless plasmas and this phenomenon
is known as {\sl{Landau Damping}} \cite{LD,JR}. Being independent of
collisions, it is fundamentally different from dissipation in ordinary
absorbing media; the collisionless dissipation does not involve an increase
of entropy and is therefore a thermodynamically reversible process.
In this phenomenon, there is a continuous exchange of energy between plasma waves 
(produced by some external perturbation) and plasma particles. As a consequence 
the self energy of plasma particles become
complex and they loose energy to plasma wave reversibly and in the very next step 
again gain energy from the plasma waves and the process of loss and gain
continues, but does not decay to other particles. This phenomenon makes the system
mechanically stable. 

The evaluation of nucleon self energy $\Sigma$ in relativistic approaches has been
restricted mainly to the real part of $\Sigma$ in either the {\sl{mean field
theory}} or the {\sl{relativistic Hartree approximation}}. In these
approximations the self-energy is real, local and energy independent. It is
completely determined once the constant effective mass and the mean constant
value of vector potential is know \cite{SW}.

However, the imaginary part of the self-energy, which in reality the
imaginary part of optical potential has been studied to lowest order by
Horowitz \cite{CJH}. Here, only the on-shell properties of
$Im(\Sigma(\varepsilon, \vec p))$ is important and is restricted to a
limited energy range. He obtained a quantity similar to the imaginary part of 
optical potential by the cancellations of large contributions coming from the
exchange of scalar and vector mesons; between scalar, time-like and
three-vector contributions which are also independently very large.

In ref. \cite{PRC41}, real and imaginary parts of nucleon self-energy was
obtained with {\sl{relativistic Dirac-Bruckner}} approach assuming
$\sigma-\omega$ meson exchange. In this paper fourth order meson-nucleon
coupling was considered and studied the off-shell behaviour of imaginary
part of the nucleon self energy.

In the reference \cite{DB} Dyson equation is used to obtain the nucleon
self energy $\Sigma^*$. In this paper the irreducible self energy is calculated 
by summing over the Dyson expansion to all order. It is shown that in the lowest 
order diagrams the irreducible self-energy contains mean field part and the
Hartree-Fock contribution. Both of them are real in nature. The dynamic part
of the self energy is expressed in terms of the four point vertex function.

However, the complex nature of nucleon self-energy as we
have obtained in this article, arises because of some kind of mixing in the
iso-spin space (the modified form of {\sl{Projection Operators}} e.g., 
$\Lambda_{pn}$ and $\Lambda_{np}$ are actually responsible for such strange
phenomena. These modified form of projection operators are again coming because 
in presence strong quantizing magnetic field, the Landau levels of protons are only 
populated, whereas neutrons behave classically) and it is at present very difficult 
to predict what will happen to these nucleons with complex energy.

\end{document}